\documentclass[useAMS,usenatbib,usegraphicx]{mn2e}

\def\l{{\ell}}
\def\lm{{\l m}}
\def\summ{\sum_{m=-\ell}^{\ell}}
\def\suml{\sum_{\ell=0}^{\infty}}
\def\alm{a_{\lm}}
\def\ylm{Y_{\lm}}
\def\cl{C_{\l}}

\title[Zonal Modes of CMB Temperature Maps]{Zonal Modes of Cosmic Microwave Background Temperature Maps}
\author[Short \& Coles]{Jo Short\thanks{Email: ShortJ1@cardiff.ac.uk (JS); Peter.Coles@astro.cf.ac.uk (PC)}
 and Peter Coles\\
 School of Physics and Astronomy, Cardiff University,
  Queens Buildings, 5 The Parade, Cardiff, CF24 3AA, United Kingdom.}

\begin{document}

\maketitle

\begin{abstract}
All-sky maps of the cosmic microwave background temperature
fluctuations are usually represented by a spherical harmonic
decomposition, involving modes labeled by their degree $l$ and order
$m$ (where $-l\leq m \leq +l$). The {\em zonal modes} (i.e those
with $m=0$) are of particular interest because they vary only with
galactic latitude; any anomalous behavior in them might therefore
be an indication of erroneous foreground substraction. We perform a
simple statistical analysis of the modes with low $l$ for sky maps
derived via different cleaning procedures from the Wilkinson
Microwave Anisotropy Probe (WMAP), and show that the zonal modes
provide a useful diagnostic of possible systematics.
\end{abstract}

\begin{keywords}
cosmology: cosmic microwave background --- cosmology: observations
--- methods: data analysis
\end{keywords}

\section{Introduction}

Observations of the temperature anisotropies of the cosmic microwave
background (CMB), particularly those from the Wilkinson Microwave
Anisotropy Probe (WMAP) \citep{b1,b3}, form the foundations of
the remarkably successful ``concordance'' cosmological model
\citep{Colesb}.  An essential ingredient of this model is the
assumption that the primordial density fluctuations that seeded the
formation of galaxies and large-scale structure were statistically
homogeneous and Gaussian. Versions of the inflation scenario based
on the idea of a single slow-rolling scalar field predict levels of
non-Gaussianity too small to be observed. On the other hand,
multi-field inflation models, and models with a non-standard kinetic
term for the inflaton, may yield larger non-Gaussian effects which
could in principle be detected in current or next-generation
observations
\citep{BMR2002,BU2002,lyth2003,dvali2004,ACMZ2004,AST2004,BKMR2004,Chen2007,BB2007,KMVW2007}.
Analysis of currently available WMAP data provide strong limits on
the level of non-Gaussianity
\citep{Komatsu2003,Spergel2007,Creminelli2007,hikage}. On the other
hand, \citet{YW2007} recently reported a detection of primordial
non-Gaussianity at greater than 99.5\% significance. Further
detailed analyses of non-Gaussianity are clearly necessary in order
to reconcile and understand the various constraints and claimed
detections.

The greatest barrier to the detection of non-Gaussianity, or other
departures from the framework of the concordance cosmological model,
is the presence of residual foreground contamination or other
systematic errors. Since our own Galaxy emits at microwave
frequencies, the emission from local foregrounds must be carefully
cleaned before a map can be obtained that is suitable for analysis.
One way of avoiding this problem is to cut out regions near the
Galactic plane where contamination is particularly severe, but this
throws away the advantage of having full coverage. It is therefore
important to produce maps that are as clean as possible over the
whole sky for many purposes. However, such cleaning is inevitably
approximate and biases are bound to occur \citep{b6,nvn,cnc3}.
Circumstantial evidence exists that may be interpreted as being due
to the presence of residual Galactic foregrounds in the WMAP data, or
some other artefact of the cleaning process \citep{cnc2,ch07}, but
in the absence of a more complete characterization of the galactic
emission the situation remains unclear.

In this {\em Article} we propose and test a simple diagnostic analysis
that offers the possibility of identifying foreground-related biases
and systematics in all-sky maps of the CMB.

\section{Zonal Modes of CMB Maps}

The statistical variation of the CMB temperature, $T(\theta,\varphi)$,
over the celestial sphere can be conveniently decomposed into
spherical harmonic modes:
\begin{equation}
T(\theta,\varphi)=\suml \summ \alm \ylm (\theta,\varphi),
\end{equation}
where the $\ylm(\theta,\varphi) $ are spherical harmonic functions,
defined in terms of the Legendre polynomials, $P_\lm$, using
\begin{equation}
\ylm(\theta,\varphi)=(-1)^m
\sqrt{\frac{(2\l+1)(\l-m)!}{4\pi(\l+m)!}}P_\lm(\cos\theta)e^{i m
\varphi},
\end{equation}
and the $\alm$ are complex coefficients which can be expressed with
$\alm=|\alm| \exp(i \Phi_\lm)$ where $\Phi_\lm$ are the phases
\citep{coles,cnc1,cnc2,cnc3,sc}. We have adopted the Condon-Shortly
phase definition in the spherical harmonic decomposition.

The spherical harmonics functions, $\ylm$, can be visualized by
considering their {\em nodal lines}, i.e. the set of points
$(\theta,\varphi)$ on the sphere where the spherical harmonics
vanish, i.e. where $\ylm(\theta,\varphi)=0$. Nodal lines are
circles, which are either in the latitude or longitude direction
with respect to the coordinate system being used; in the case of CMB
maps this is usually the galactic coordinate system. The number of
nodal lines of each type is determined by the number of zeros of
$\ylm$ in the latitudinal and longitudinal directions independently.
The associated Legendre functions $P_\lm$ possess $\l-|m|$ zeros in
the latitude direction, whereas the trigonometric $\sin$ and $\cos$
functions possess $2|m|$ zeros in the longitude direction. Two
specific orders $m$ are of particular interest at a given $l$ in the
context of this work: these denote the {\em zonal} modes, with
$m=0$, and the {\em sectoral} modes, with $m=\l$. In the former case
there are no zero-crossings in the longitude direction, so contours
of equal temperature run parallel to latitude lines; in the latter
the contours run parallel to longitude lines. Examples are shown in
Figure 1. In the intermediate cases with $0<m<\l$ there are
zero-crossings in both directions, giving rise to a patchwork
appearance; these are usually called {\em tesseral} modes.
\begin{figure}
  \centering
  \includegraphics[width = 84mm]{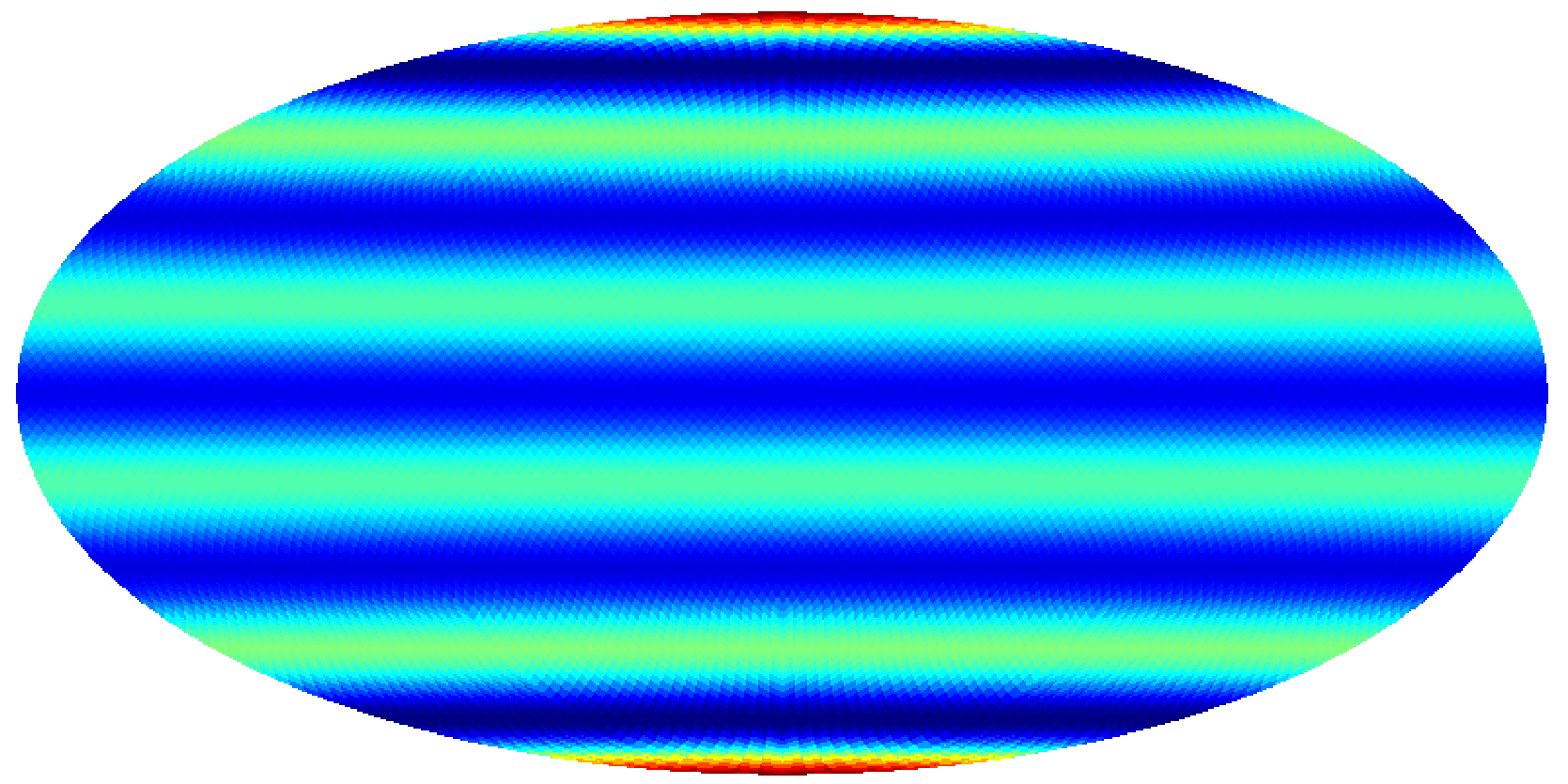}
  \includegraphics[width = 84mm]{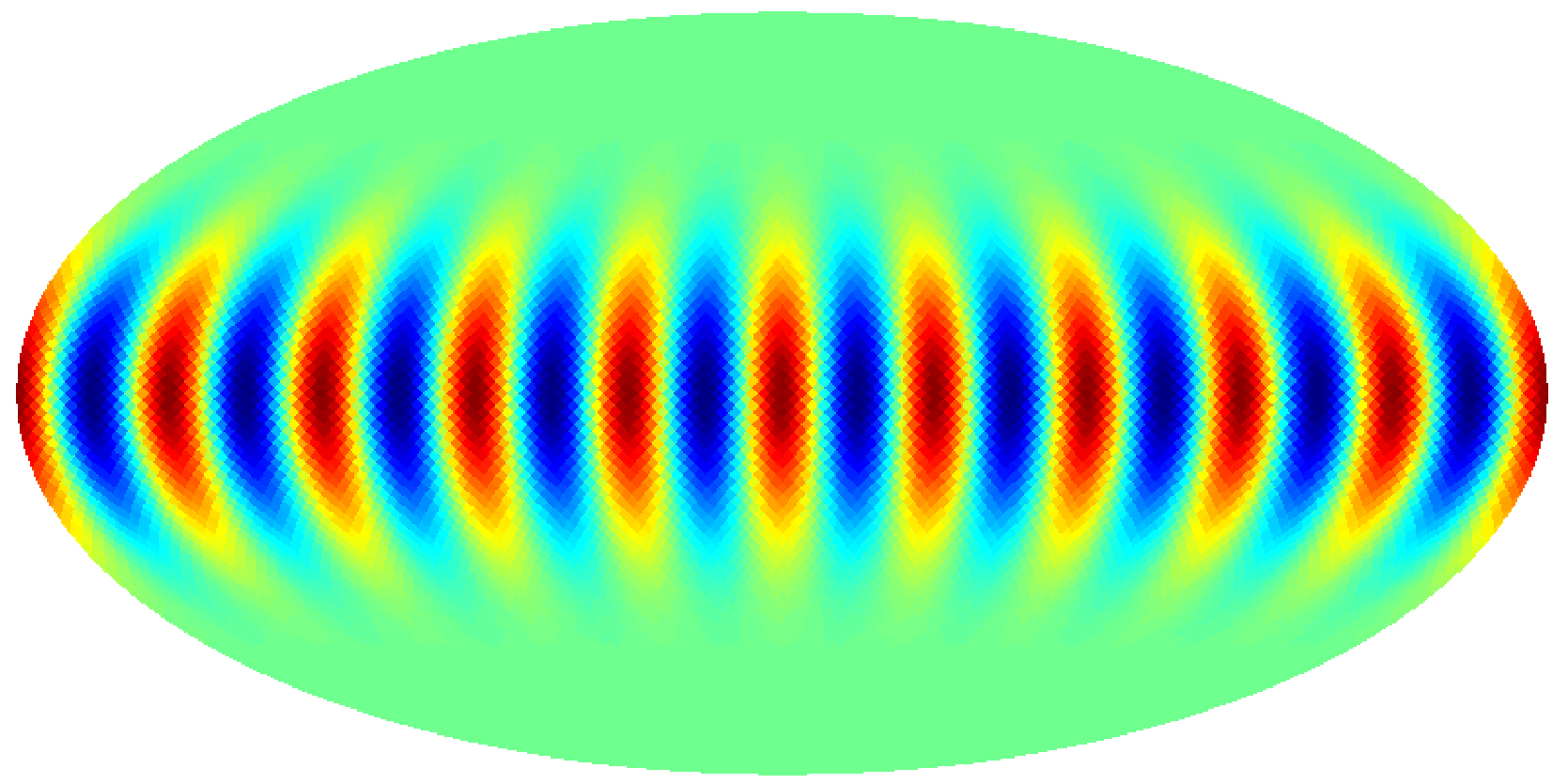}
  \caption{Illustrative examples of zonal and sectoral modes for $\l=10$; the first example is the
  zonal mode with $m=0$ and the second is the sectoral mode with $m=10$.}
  \label{fig1}
\end{figure}

Statistically isotropic Gaussian random CMB temperature fluctuations
on a sphere, of the type that result from the simplest versions of
the inflationary paradigm, possess spherical harmonic coefficients
($\alm$) whose real and imaginary parts are mutually independent and
both Gaussian \citep{be,coles}. The statistical
properties of the fluctuations are then completely specified by the
angular power spectrum, $\cl$, where
\begin{equation}
\langle  a^{}_{\l^{ } m^{ }} a^{*}_{\l^{'} m^{'}} \rangle = \cl \;
\delta_{\l^{ } \l^{'}} \delta_{m^{} m^{'}}.
\end{equation}
Since $T$ is always real, the complex vectors of the $\alm$ on the
Argand plane for $m<0$ are mirror images of those with $m>0$ with
respect to $x$ axis for even $m$, and with respect to $y$ axis for
odd $m$. Writing \begin{equation} \alm = x_\lm + i y_\lm,
\end{equation}
we note that the variances of the real and imaginary parts of $\alm$
for $m>0$ are equal
\begin{equation}
\sigma^2(x_{\lm})=\sigma^2(y_{\lm})\equiv
\sigma^2_{\l}=\frac{1}{2}\cl,
\end{equation}
which depends only on $\l$. The distributions of $x_{\lm}$ and
$y_{\lm}$ are independent Gaussians with these variances and zero
mean. The amplitudes $|\alm|$ for these modes therefore have a Rayleigh
distribution with random phases \citep{be,sc}. For $m=0$, the imaginary part of $\alm$
must be zero, so this mode always has zero phase $\Phi_{\lm}$. This is because
the phase relates to the variation around the polar axis only.
However, the distribution $x_{\lm}$ for $m=0$ should be equal to
that of the other real parts at a given $\l$, namely a Gaussian with
zero mean and stated variance.

\section{Statistical Analysis}

\subsection{Preamble}

Statistical analysis of the spherical harmonics of CMB maps usually
involves using all modes in an equivalent manner. However, since the
zonal modes in particular have such a special relationship to the
galactic coordinate system, it is worth looking at their properties
independently to see whether they hold any clues to possible
residual contamination aligned with the Galactic plane.

In order to construct a test which involved the smallest possible
number of assumptions, and in particular avoided the need to make
estimates of the power-spectrum $\cl$ along the way, we focused on
the modes with maximum or minimum amplitude at a given $\l$. If all
modes are statistically equivalent then the different orders $m$ at
a given $\l$ are equally likely to furnish the maximum (or minimum)
amplitude $|\alm|$. A preference for modes with $m=0$ to display the
maximum (or minimum) amplitude might therefore  be plausibly
interpreted as evidence that the zonal modes are either contaminated
with residual foreground, or that foregrounds have been excessively
subtracted. Both of these possibilities are supported by other
evidence \citep{cnc3,ch07,nvn}.

\subsection{Data}

The maps analyzed in this paper are the 1-year \citep{b1}, 3-year
\citep{b2}, and 5-year \citep{b3} Internal Linear Combination (ILC)
maps from the WMAP team, the 1 and 3 year maps from \cite{b4} and
the harmonic ILC map by \cite{b5} (hereafter ILC1, ILC3, ILC5, TOH1,
TOH3 and HILC). We apply three tests to these maps, which are
described in detail below.

\subsection{ Extremal Mode Counts}

For our first test, we analyze the value of $m$ associated with the
largest (or smallest) amplitude $|\alm|$ for a given $\l$ in the
ranges $[0,10]$ and $[0,20]$. Since the $\alm$ for all orders $m$ at
a given $\l$ should have the same variance in the null hypothesis
(of a stationary Gaussian random field), the minimum or maximum
value of $|\alm|$ should not occur preferentially at any particular
value of $m$. We use a very straightforward method to establish
whether this is the case. For each map we simply count the number of
occurrences (i.e. numbers of separate degrees $\l$ within the range
analyzed) for which the minimum or maximum value amplitude occurs at
$m=0$ (the zonal mode) or $m=\l$ (the sectoral mode). These counts
are recorded in Table \ref{TABnum}. What we have done is to
identify, at each degree $\l$, the value of $m$ which has the
minimum (or maximum) value of $|\alm|$. If the minimum is at $m=0$
then this contributes to the count in Column 3 of the table; if the
maximum is $m=0$ it contributes to the count in Column 4; likewise
occurrences of extrema at $m=\l$ contribute to the counts in the
following two columns. For results where the maximum $\l$ = 10 the
counts are out of a total of 11 and for maximum $\l$ = 20 they are
out of a total of 21.

\begin{table}
  \begin{tabular}{|c|c|c||c|c|c|}
    \hline \hline
    Map &  $\l_{\rm max}$ & $m_{\rm min}=0$ & $m_{\rm max}=0$  & $m_{\rm min}=\l$ & $m_{\rm max}=\l$ \\ \hline
    ILC1     & 10 & 6    & 3  &  3    &  4    \\
    ILC3     & 10 & 7    & 2  &  2    &  4    \\
    ILC5     & 10 & 7    & 2  &  2    &  4    \\
    TOH1     & 10 & 6    & 2  &  3    &  4    \\
    TOH3     & 10 & 8    & 2  &  2    &  4    \\
    HILC     & 10 & 8    & 2  &  1    &  5    \\ \hline
    ILC1     & 20 & 9    & 3  &  3    &  6    \\
    ILC3     & 20 & 9    & 2  &  2    &  5    \\
    ILC5     & 20 & 8    & 2  &  2    &  4    \\
    TOH1     & 20 & 8    & 3  &  3    &  5    \\
    TOH3     & 20 & 11   & 2  &  2    &  5    \\
    HILC     & 20 & 10   & 2  &  1    &  6    \\ \hline
    \hline
  \end{tabular}
  \caption{This table shows, for the various cleaned all-sky maps described in the text, the
  maximum value of the degree $\l$ considered, and, in the following four columns, the number of
  times (i.e. number of values of $\l$) for which the minimum (or
  maximum) amplitude $\alm$ is at $m=0$ (zonal mode) or $m=\l$
  (sectoral mode). For example, the $m_{\rm min}=0$ and max $\l$ = 20 result for TOH3
means that 11 out of 21
  different values of the degree $\l \in [0,20]$ have the lowest value of the amplitude $|\alm|$ at $m=0$.}
  \label{TABnum}
\end{table}

We have restricted our analysis to low $\l$ modes partly to keep the
computational cost of the simulations down but mainly because the
high $\l$ modes are known not to be clean anyway so it would not
reveal anything interesting to find zonal or sectoral anomalies
among them. We have also given results for the sectoral modes for
comparison, but there are no significant anomalies associated with
them and we shall not discuss them further in this paper.

\subsection{Significance Levels}

It is not a trivial matter to calculate significance levels
analytically for this test because the number of available orders
$m$ increases with $\l$. It is obviously much more probable that the
minimum amplitude is at $m=0$ for $\l=2$ than for $\l=20$ under the
null hypothesis. Assessing the significance of the number of
occurrences of zonal or sectoral extrema involves a messy exercise
in combinatorics. However, we can finesse this difficulty by instead
comparing the actual maps with simulations constructed according to
the Gaussian assumption described in Section 2.

It is also possible that the cleaning process used to remove
foreground contamination from the raw observations might in any case
introduce some sort of bias into the statistical distribution of
amplitudes or induce correlations between different modes. To
circumvent this difficulty, as well as the one noted in the previous
paragraph, we therefore base the confidence levels on our test on a
set of simulations performed by \cite{b6}, which takes ``raw'' sky
maps, generated assuming Gaussian fluctuations, adds simulated
foregrounds and then recovers the signal using the ILC methodology.
The simulated maps we use therefore already take into account any
``artificial'' correlations that the ILC process may generate. These
are particularly useful as they allow us to assess whether any
anomalies we do actually find must be above the level known to be
introduced by the ILC cleaning process.

In order to calculate significance levels of the results in Table
\ref{TABnum} we used an ensemble of $N=10000$ independent Monte
Carlo realizations of Gaussian skies, contaminated with foreground
and then cleaned according to the ILC prescription as described
above. We use these simulations to construct empirical distributions
of the count statistics displayed in the previous table and from
these we compute the empirical significance levels shown in Table
\ref{TABsig}.

The corresponding probabilities of the results in Table \ref{TABsig}
were calculated using the Monte Carlo simulations detailed above by
counting the number of occurrences of each configuration shown in
Table \ref{TABnum}. Note that since we are using $N=10000$
independent simulations we expect the results to be affected by
Poisson fluctuations at the level of order $\sqrt{N}$, which means
we expect the probabilities in Table \ref{TABsig} only to be
accurate to about 1\% or so.

For the case of zonal maxima, it is clear that there no significant
results above 2$\sigma$ (i.e. the 95\% level), but the zonal minima
do show a significant result in the TOH3 and HILC maps for $\l<10$.
It is also interesting to note that the TOH3 map gives a higher
significance level than the TOH1 map, as does the ILC5 map compared
to the corresponding 1 yr map ILC1.
\begin{table}
 \begin{center}
  \begin{tabular}{|c|c|c||c|c|c|}
    \hline \hline
    Map &  $\l_{\rm max}$ & $m_{\rm min}=0$ & $m_{\rm max}=0$ & $m_{\rm min}=\l$ & $m_{\rm max}=\l$ \\
    \hline
    ILC1&   10&       75.2 &   15.4 &             64.2 &             80.4\\
    ILC3&   10&       91.7 &    0.0 &             25.5 &             80.4\\
    ILC5&   10&       91.7 &    0.0 &             25.5 &             80.4\\
    TOH1&   10&       75.2 &    0.0 &             64.2 &             80.4\\
    TOH3&   10&       98.2 &    0.0 &             25.5 &             80.4\\
    HILC&   10&       98.2 &    0.0 &              0.0 &             94.9\\ \hline
    ILC1&   20&       88.0 &    6.3 &             46.8 &             92.8\\
    ILC3&   20&       88.0 &    0.0 &             15.6 &             80.1\\
    ILC5&   20&       74.1 &    0.0 &             15.6 &             57.1\\
    TOH1&   20&       74.1 &    6.3 &             46.8 &             80.1\\
    TOH3&   20&       98.4 &    0.0 &             15.6 &             80.1\\
    HILC&   20&       95.0 &    0.0 &              0.0 &             92.8\\ \hline
    \hline
  \end{tabular}
  \caption{Monte Carlo estimates of the probabilities of the extremal mode counts,  i.e
   occurrences of $m_{\rm ext} = 0$ or $\l$, for the various CMB maps shown in Table
   \ref{TABnum}. These are computed by forming the empirical
   distribution of the counts over an ensemble of simulated skies
   and counting what fraction of the ensemble gives the results
   obtained for the real maps. For example, in the case of the $m_{\rm min}=0$ and max $\l$ = 20
result for TOH3, we find   that, out of 10000 simulations, 9844 have
{\em less than} 11 (from Table \ref{TABnum})
  occurrences of minimum amplitudes at $m=0$. Given the probable sampling accuracy of around one
  percent,
  we have rounded the results. Note that these are discrete distributions, so the zero
  percentages do not necessarily indicate cases of exceptional significance, just that it is not possible to have less than the
  observed number of occurrences. In other words, we should treat this as a one-sided statistical test. }
  \label{TABsig}
\end{center}
\end{table}

It is worth stressing that each column represents a single
statistical test using all the information contained in the modes
with degree $\l$ in the range analyzed, and therefore represent
genuine {\em a priori} significance levels. For example, in the case
of the $m_{\rm min}=0$ and max $\l=20$ result for TOH3 described in
the caption for Table \ref{TABsig}, the probability that a genuinely
Gaussian CMB sky processed in the way we described above, would
produce as many zonal minima as observed is only $(100-98.4) =
1.6$\%.

\subsection{Mode Variances}

We performed a second investigation in order to locate the source of
the apparent deficit in the zonal modes. The results above suggest
that the amplitudes when $m=0$ are low compared
 to simulations, so this test considers whether this is because the variance of the
 amplitudes is lower for $m=0$ than for all other m. Since there is no imaginary
part to the amplitude when $m=0$, just the real parts of the $\alm$
(i.e. the $ x_{\lm}$) are considered. The variances of the $
x_{\l0}$ for the ILC5 and TOH3 maps are calculated and the
corresponding probabilities are calculated by comparing these values
to the variances of the $ x_{\l0}$ from the simulations. The results
in Table \ref{TABvar} provide evidence that the variance of the
$x_{\l0}$ in the ILC5 and TOH3 maps is significantly smaller than in
the simulations. For comparison, the same calculation is also
applied `for all m' and `for all m not equal to 0'. The results for
all $m$ show that the variance in the TOH3 and ILC5 maps is lower
than in the simulations, which is consistent with the previously
reported low variance in the combined Q + V + W map by \cite{b7}.
This leads to the further question of whether the zonal mode
amplitudes are entirely responsible for this low variance. Table
\ref{TABvar} shows that by removing the $m = 0$ amplitudes the
number of simulated maps with a variance greater than the test map
falls dramatically, suggesting that indeed the low variance of the
$m=0$ amplitudes does has a notable affect on the overall variance
of the amplitudes.

\begin{table}
\begin{center}
  \begin{tabular}{|c|c|c|c|}
    \hline \hline
    Map & $m = 0$ & $\forall m$ & $m\ne 0$    \\ \hline
    ILC5  &  97.1  & 93.2 &    69.0             \\
    TOH3  &  97.6  & 93.8 &    69.2            \\  \hline
    \hline
  \end{tabular}
  \caption{Percentage of simulations, for given $m$, where the
  variance of the $x_{\lm}$ is greater than the specified map.
  For example take the TOH3 result for m=0. We found that, out of 10000 simulations,
  9757 had a variance of the $x_{\l 0}$  greater that the variance of the $x_{\l 0}$ in the TOH3 map.
  Since the Poisson fluctuations are of order 1\% we have rounded the resulting percentage.}
  \label{TABvar}
\end{center}
\end{table}

\begin{figure}
  \centering
  \includegraphics[width = 84mm]{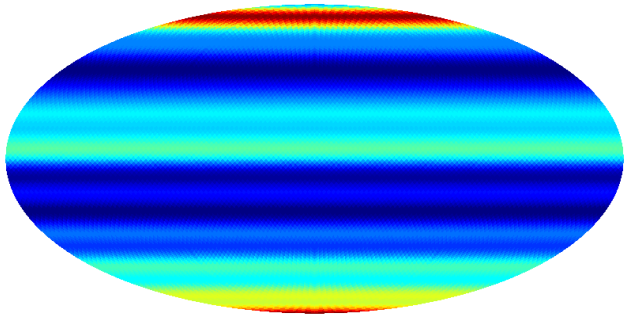}
  \caption{Reconstructed CMB map for WMAP ILC 5 yr map using only the $m=0$ modes for $\l = [0,20]$
  i.e. this map is the WMAP ILC 5 yr map which has had all but the $|a_{\l 0}|$ set to zero.
  The colour scale is marked in milliKelvin.}
  \label{fig2}
\end{figure}

\subsection{Map Structure}

It is interesting to reconstruct what the CMB sky would look like if
it only contained the zonal modes, as these are the ones that appear
to have anomalous properties; the result in Figure \ref{fig2} shows
that the deepest minima are just below and slightly above the
Galactic plane. For our third test we considered whether these
minima (and also the maxima of the map) are of abnormal amplitude
compared to the simulated maps; this test is done in pixel-space
rather than using the $\alm$. Because the resolution of the maps is
quite low, we restricted our analysis to the parts of the map that
are well-sampled (i.e. we neglected the Galactic pole areas). Monte
Carlo simulations were run to estimate the probability of a maximum
or minimum as seen in the map. The resulting probabilities are shown
in Table \ref{TABcol}. Similarly to \cite{LW2004}, who find that the
hot and cold spots of the separate WMAP frequency maps are not hot
and cold enough, we find that the maxima (and minima) of the zonal
maps are not as high (and low) as expected compared to the
simulations.
\begin{table}
\begin{center}
  \begin{tabular}{|c|c|c|}
    \hline \hline
    Map      &  minimum &  maximum \\ \hline
    ILC1     &  73.0   &   88.1    \\
    ILC3     &  82.1   &   98.9    \\
    ILC5     &  81.2   &   99.0    \\
    TOH1     &  77.9   &   95.0    \\
    TOH3     &  95.2   &   98.6    \\
    HILC     &  85.4   &   97.9    \\ \hline
    \hline
  \end{tabular}
  \caption{Probability of the observed maxima and minima in zonal maps, as derived from Monte Carlo simulations.
  For example take the results for ILC5: out of 10000 simulations,
  8124 have a minimum `temperature' in the $m=0$ only map which is {\em less than} that observed in the ILC5 $m=0$ map
  (see Figure \ref{fig2}).
  Similarly, out of 10000 simulations, 9902 have a maximum `temperature' in the $m=0$  map which is
  {\em greater than} that observed in the ILC5 $m=0$ map.}
  \label{TABcol}
\end{center}
\end{table}
The trend reinforced again in this table seems to be that the
anomalous result becomes increasingly significant as the maps
supposedly improve.

\section{Discussion}
We have presented a simple statistical analysis based on properties
of the zonal modes of cosmic microwave background maps, i.e. those
aligned parallel to the Galactic plane. An application of the test
to various cleaned CMB maps gives interesting results. At the 95 per
cent level, no significant anomalies appear in the WMAP ILC maps
\citep{b1,b2,b3} but there seems to be a significant tendency in
some other maps \citep{b4,b5} to have zonal modes with
systematically lower amplitudes than would be expected in the
concordance model. Intriguingly, the maps that provide the most
significant departures from the behavior expected under the null
hypothesis are those based on later issues of the WMAP data.  We
have also showed that the low variance of the maps is more
significant where you consider only the zonal amplitudes. This shows
that zonal modes have a notable contribution towards this low
variance. Finally, we considered the distribution of maxima and
minima in zonal maps which again reinforced the earlier finding that
the anomalies increase with later data releases. The question of
whether this is due to increasing over subtraction, or whether
decreasing noise is revealing an underlying anomaly such as a
systematic problem, cannot be distinguished with current data.

Of course the maps themselves are {\em} not statistically
independent. Indeed, if the cleaning processes involved were perfect
then they would all be identical. The different results we have
found for the different maps are attributable to the nature of the
zonal modes and their extra sensitivity to the structures associated
with the Galactic Plane. The appearance of significant anomalies in
some maps rather than others is not a statistical fluke but is clear
evidence that some cleaning methods leave artifacts in the
distribution of mode amplitudes.

It must be noted that the probabilities we quote of around 98 to 99
percent are not overwhelming, so the results we have obtained are
indicative rather than decisive. This is not surprising, given the
relatively small number of modes we used. However, we repeated the
analysis for a coordinate system aligned with the Ecliptic, rather
than Galactic plane, and found no significant results at all. This
lends further credence to our interpretation of the outcome of our
analysis in terms of an effect related to over-subtraction of
Galactic emission \citep{cnc3,nvn}, which becomes increasingly
pronounced with each data release. A more definitive result will
have to wait until more detailed foreground subtraction can be
attempted, such as will be the case with the {\em Planck} satellite.

\section*{Acknowledgments}
Jo Short receives support from a Science \& Technology
Facilities Council (STFC) doctoral training grant. We would like to
thank Pavel Naselsky for interesting discussions. We gratefully
acknowledge the use of the HEALPIX package \citep{heal}.


\begin{thebibliography}{99}
\bibitem[\protect\citeauthoryear{Alishahiha et al.}{2004}]{AST2004}
  Alishahiha M., Silverstein  E., Tong D., 2004,
  Phys. Rev. D., 70, 123505
  \bibitem[\protect\citeauthoryear{Arkami-Hamed et al.}{2004}]{ACMZ2004}
  Arkani-Hamed N., Creminelli P., Mukohyama S., Zaldarriaga M., 2004,
  JCAP, 4, 1
\bibitem[\protect\citeauthoryear{Bartolo et al.}{2002}]{BMR2002}
  Bartolo N., Matarrese S., Riotto A., 2002, Phys. Rev. D., 65, 103505
  \bibitem[\protect\citeauthoryear{Bartolo et al.}{2004}]{BKMR2004}
  Bartolo N., Komatsu E., Matarrese S., Riotto A., 2004, Phys. Rept., 402, 103
  \bibitem[\protect\citeauthoryear{Battefeld \& Battefeld}{2007}]{BB2007}
  Battefeld D., Battefeld T., 2007, JCAP, 5, 12
\bibitem[\protect\citeauthoryear{Bennett et al.}{2003}]{b1} Bennett C. L.  et al., 2003, ApJS, 148,
1
\bibitem[\protect\citeauthoryear{Bernardeau \& Uzan}{2002}]{BU2002}
  Bernardeau F., Uzan J.-P., 2002,  Phys. Rev. D., 66, 103506
\bibitem[\protect\citeauthoryear{Bond \& Efstathiou}{1987}]{be}
Bond J. R., Efstathiou G., 1987, MNRAS, 226, 655
\bibitem[\protect\citeauthoryear{Chen et al.}{2007}]{Chen2007}
  Chen X., Easther R., Lim E. A., 2007, JCAP, 6, 23
\bibitem[\protect\citeauthoryear{Chiang, Naselsky \& Coles}{2004}]{cnc1}
Chiang L.-Y., Naselsky P. D., Coles P., 2004, ApJ, 602, L1
\bibitem[\protect\citeauthoryear{Chiang, Naselsky \& Coles}{2007}]{cnc2}
Chiang L.-Y., Naselsky P. D., Coles P., 2007, ApJ, 664, 8
\bibitem[\protect\citeauthoryear{Chiang, Naselsky \& Coles}{2009}]{cnc3}
Chiang L.-Y., Naselsky P. D., Coles P., 2009, ApJ, 694, 339
\bibitem[\protect\citeauthoryear{Chiang et al.}{2007}]{ch07}
Chiang L.-Y., Coles P., Naselsky P. D., Olesen P., 2007, JCAP, 1, 21
\bibitem[\protect\citeauthoryear{Coles} {2005}]{Colesb} Coles P., 2005, Nat, 433, 248
\bibitem[\protect\citeauthoryear{Coles et al.}{2004}]{coles}
 Coles P., Dineen P., Earl J., Wright D., 2004, MNRAS, 350, 989
\bibitem[\protect\citeauthoryear{Creminelli et al.}{2007}]{Creminelli2007}
  Creminelli P., Senatore L., Zaldarriaga M., \&
  Tegmark M., 2007, JCAP, 3, 5
 \bibitem[\protect\citeauthoryear{Dvali et al.}{2004}]{dvali2004}
  Dvali G., Gruzinov A., Zaldarriaga M., 2004,
  Phys. Rev. D., 69, 083505
\bibitem[\protect\citeauthoryear{Eriksen et al.}{2005}]{b6} Eriksen H. K.  et al., 2005, astro-ph/0508196
\bibitem[\protect\citeauthoryear{G\'{o}rski et al.}{2005}]{heal} G\'{o}rski K. M., Hivon E., Banday A. J., Wandelt B. D.,
Hansen F. K., Reinecke E. M., Bartelmann M., 2005, ApJ, 622, 759
\bibitem[\protect\citeauthoryear{Hikage et al.}{2008}]{hikage}
Hikage C., Matsubara T., Coles P., Liguori M., Hansen H. K.,
Matarrese S., 2008, MNRAS, 389, 1439
\bibitem[\protect\citeauthoryear{Hinshaw et al.}{2009}]{b3} Hinshaw G.  et al., 2009, ApJS, 180, 225
\bibitem[\protect\citeauthoryear{Jarosik et al.}{2007}]{b2} Jarosik N.  et al., 2007, ApJS, 170, 263
\bibitem[\protect\citeauthoryear{Kim et al.}{2008}]{b5} Kim J. et
al., 2008, Phys. Rev. D, 77, 103002
\bibitem[\protect\citeauthoryear{Komatsu et al.}{2003}]{Komatsu2003}
  Komatsu E. et al., 2003, ApJS, 148, 119
\bibitem[\protect\citeauthoryear{Koyama et al.}{2007}]{KMVW2007}
  Koyama K., Mizuno S., Vernizzi F., Wands D., 2007, JCAP, 11, 24
\bibitem[\protect\citeauthoryear{Larson \& Wandelt}{2004}]{LW2004}
  Larson D. L., Wandelt B. D., 2004, ApJ, 613, L85-88
\bibitem[\protect\citeauthoryear{Lyth et al.}{2003}]{lyth2003}
  Lyth D. H., Ungarelli C., Wands D., 2003,
  Phys. Rev. D., 67, 023503
\bibitem[\protect\citeauthoryear{Monteserin et al.}{2008}]{b7} Monteserin C.  et al., 2008, MNRAS, 387, 209
\bibitem[\protect\citeauthoryear{Naselsky, Verkhodanov \&
Nielsen}{2008}]{nvn} Naselsky P. D., Verkhodanov O. V., Nielsen M.
T. B., 2008, AstBu,  63, 216
\bibitem[\protect\citeauthoryear{Spergel et al.}{2007}]{Spergel2007}
  Spergel D. N. et al., 2007, ApJS, 170, 377
\bibitem[\protect\citeauthoryear{Stannard \& Coles}{2005}]{sc}
Stannard A., Coles P., 2005, MNRAS, 364, 929
\bibitem[\protect\citeauthoryear{Tegmark et al.}{2003}]{b4} Tegmark M. et al., 2003, Phys. Rev. D, 68, 123523
\bibitem[\protect\citeauthoryear{Yadav \& Wandelt}{2008}]{YW2007}
  Yadav A. P. S., Wandelt B. D., 2008, Phys. Rev. Lett., 100, 181301
\end{thebibliography}
\end{document}